\documentclass[12pt]{extarticle}
\usepackage[utf8]{inputenc}
\usepackage{cite}
\RequirePackage{bm}
\usepackage{amsgen,amsmath,amstext,amsbsy,amsopn,amssymb}
\usepackage{algorithm}
\usepackage{algorithmic}
\usepackage{pifont}
\newcommand{\norm}[1]{\left\lVert#1\right\rVert}
\usepackage{amsthm}
\usepackage{graphicx}

\usepackage{newclude}

\newtheorem{theorem}{Theorem}

\newtheorem{proposition}{Proposition}
\newtheorem{lemma}{Lemma}

\title{The AUGUST Two-Sample Test: Powerful, Interpretable, and Fast}
\author{Benjamin Brown\footnote{Benjamin Brown is a graduate student (Email: brownb1@live.unc.edu), Department of Statistics and Operations Research, University of North Carolina at Chapel Hill.}, Kai Zhang\footnote{Kai Zhang is Associate Professor (Email: zhangk@email.unc.edu), Department of Statistics and Operations Research, University of North Carolina at Chapel Hill. Zhang's research was partially supported by NSF DMS-1613112, NSF IIS-1633212, NSF DMS-1916237.}}

\begin{document}

\maketitle

\begin{abstract}

Two-sample testing is a fundamental problem in statistics, and many famous two-sample tests are designed to be fully non-parametric. These existing methods perform well with location and scale shifts but are less robust when faced with more exotic classes of alternatives, and rejections from these tests can be difficult to interpret. Here, we propose a new univariate non-parametric two-sample test, AUGUST, designed to improve on these aspects. AUGUST tests for inequality in distribution up to a predetermined resolution using symmetry statistics from binary expansion. The AUGUST statistic is exactly distribution-free and has a well-understood asymptotic distribution, permitting fast $p$-value computation. In empirical studies, we show that AUGUST has power comparable to that of the best existing methods in every context, as well as greater power in some circumstances. We illustrate the clear interpretability of AUGUST on NBA shooting data.

\end{abstract}

\section{Introduction}

Two-sample tests are one of the most frequently used methods for statistical inference. Across disciplines, we encounter the same setup: two samples $\bm{X}$ and $\bm{Y}$ correspond to two different conditions, and we want to decide if the condition affects the data in some way. Does a drug improve patient outcomes compared to placebo? Are test subjects' reaction times more variable in one context than another? In general, we want to test the null hypothesis that $\bm{X}$ and $\bm{Y}$ come from the same distribution. For now, we will assume $\bm{X}$ and $\bm{Y}$ are univariate, though we discuss multivariate extensions in Section \ref{sec:multi}.

Often, we can assume no specific knowledge of the distributions $F$ and $G$ used to generate $\bm{X}$ and $\bm{Y}$. In this fully non-parametric setting, the problem above is difficult to approach. Some well-known rank-based tests are designed for this context: the Mann-Whitney $U$-test \cite{mann1947test} looks for a location shift between $F$ and $G$, while the Lepage \cite{lepage1971combination} and Cucconi \cite{cucconi1968nuovo} tests address both location and scale. However, even if $F$ and $G$ are not location or scale shifted, differences may hide in the distributions' modality, skewness, local shape, and so forth.

In order to detect more elusive distributional differences, a number of non-parametric tests exist which make no assumptions about how $F$ may differ from $G$. For instance, the Kolmogorov-Smirnov statistic \cite{kolmogorov1933sulla} is the maximal vertical distance between the samples' empirical CDF curves. The Cramer-von Mises statistic \cite{cramer1928composition} sums the squared vertical distance between ECDFs at every point in the combined sample; Anderson-Darling \cite{anderson1952asymptotic} does the same while weighting each vertical distance by an estimate of its variance. Instead of a discrete sum over points in the sample, a test using Wasserstein distance \cite{dobrushin1970prescribing} uses the total area between ECDF curves. The recent DTS test \cite{dowd2020new} somewhat combines Anderson-Darling and Wasserstein, using as its statistic a variance-weighted area between ECDF curves.

Each of these non-parametric tests uses some variety of distance between the ECDFs of $\bm{X}$ and $\bm{Y}$. This approach is logical, as convergence in distribution of probability measures is equivalent to convergence of the associated CDFs in the Kolmogorov-Smirnov metric, provided that the limiting CDF is continuous. Moreover, the Glivenko-Cantelli theorem indicates that ECDFs are uniformly good approximations for their theoretical counterparts.

Yet, the ECDF-based methods do have some drawbacks. The performance of these tests is highly sensitive to 
\begin{enumerate}
\item[(1)] the type of alternative and
\item[(2)] the precise shape of the distributions, independently of the difference between them.
\end{enumerate} 
As an example of (1), these methods can excel at detecting location and scale shifts, but will struggle to catch bimodality when mean and variance are held constant. As an example of (2), we show in Section \ref{sec:performance} that the relative performance of these tests at detecting a location shift can be exactly \textit{reversed} by a suitable choice of distribution family. One would instead prefer power against location alternatives to be nearly independent of family.

One final drawback pertains to ease of use rather than power. For these methods, ECDF distance measures distributional difference as a scalar quantity. If a test rejects the null, no framework is available to determine exactly why rejection occurred. Even looking at ECDF line graphs and histograms of $\bm{X}$ and $\bm{Y}$, it can be difficult to specify which irregularities account for the rejection, as well as how much each irregularity contributes.

Here, we introduce a new non-parametric two-sample test, AUGUST -- AUGmented cdf for Uniform Statistic Transformation -- which tests for equality of distribution up to a predetermined resolution $d$. AUGUST explicitly tests for multiple orthogonal sources of distributional inequality, giving AUGUST power against a wide range of alternatives. When AUGUST rejects the null, this decomposition into orthogonal signals allows for unambiguous interpretation of exactly how equality in distribution between $\bm{X}$ and $\bm{Y}$ has failed. Moreover, the AUGUST statistic is distribution-free in finite samples, allowing for efficient computation of $p$-values via null simulation. 

In Section \ref{sec:insights}, we motivate the AUGUST test by a discussion of ECDF transformations. We show that a test of $F = G$ can be reduced to a test of the uniformity of a particular collection of ECDF-transformed variables. Consider the following transformation: for each point $x$ in $\bm{X}$, what fraction of the $\bm{Y}$ sample is less than $x$? Intuitively speaking, if these fractional values are not uniformly distributed in $[0, 1]$, then the distribution of $\bm{X}$ is a poor fit for $\bm{Y}$, and we should reject $H_0: F = G$. We build on this idea with an approach inspired by resampling.

As a central principle of AUGUST, we use the BET framework introduced in Zhang \cite{zhang2019bet}. Since we wish to test the uniformity of some collection in $[0, 1]$, we can partition $[0, 1]$ into intervals of width $1/2^d$ and record cell probabilities for each of the $2^d$ intervals, where $d$ is a fixed resolution level. Via the Hadamard transform, we map this vector of cell probabilities to a vector of symmetry statistics, thought of as a transformation from the physical domain to the frequency domain. It turns out that operating in the frequency domain simplifies the process of creating a powerful test for uniformity. Moreover, the symmetry statistics can be thought of as detecting uncorrelated sources of non-uniformity at a binary depth $d$. As a result, each symmetry statistic has a clear interpretation as to the way in which non-uniformity fails. In the context of a two-sample test, this interpretation tells us how $F \neq G$.

In Section \ref{sec:method}, we formalize the procedure for the AUGUST test and analyze its running time. Given a total sample size $N$, we prove that the AUGUST statistic can be computed in $O(N\log N)$ elementary operations, and we provide an algorithm that achieves this time complexity. Via simulation, we compare the running time of AUGUST to that of classical methods. In addition, we demonstrate how the AUGUST test can be naturally extended to a non-parametric two-sample test on multivariate data using elliptical cells.

In Section \ref{sec:theoretical}, we show that the AUGUST statistic can be written as a continuous function of a two-sample $U$-statistic. We use this fact to derive the asymptotic distribution of the AUGUST statistic, allowing faster $p$-value computation in a large-sample setting and providing a framework for power analysis under any predetermined alternative. 

In Section \ref{sec:performance}, we use simulation studies to compare the power of the AUGUST test to that of other well-known non-parametric two-sample tests. We find that AUGUST has power close to that of the best existing methods in every context, as well as greater power in some circumstances. For example, AUGUST outperforms all other tests considered at detecting unimodality versus multimodality. 

We also examine the empirical power of our multivariate extension to the AUGUST test. There are very many existing non-parametric multivariate methods, such as those of Weiss \cite{weiss1960two}; Bickel \cite{bickel1969distribution}; Baumgartner, Weiss, and Schindler \cite{baumgartner1998nonparametric}; Hettmansperger, M\"ott\"onen and Oja \cite{hettmansperger1998affine}; Hall and Tajvidi \cite{hall2002permutation}; Rousson \cite{rousson2002distribution}; Baringhaus and Franz \cite{baringhaus2004new}; Aslan and Zech \cite{aslan2005new}; Eric, Bach, and Harchaoui \cite{harchaoui2007testing}; Oja \cite{oja2010multivariate}; Gretton et al. \cite{gretton2012kernel}; Sz\'ekely and Rizzo \cite{szekely2013energy}; Biswas et al. \cite{biswas2014distribution}; Biswas and Ghosh \cite{biswas2014nonparametric}; Chwialkowski et al. \cite{chwialkowski2015fast}; Lopez-Paz and Oquab \cite{lopez2016revisiting}; Li \cite{li2018asymptotic}; Pan et. al. \cite{pan2018ball}; and Song and Chen \cite{song2020generalized}. Of recent note are the graph-based two-sample tests, including Friedman and Rafsky \cite{friedman1979multivariate}; Schilling \cite{schilling1986multivariate}; Henze \cite{henze1988multivariate}; Liu and Singh \cite{liu1993quality}; Rosenbaum \cite{rosenbaum2005exact}; Chen and Friedman \cite{chen2017new}; and Chen, Chen, and Su \cite{chen2018weighted}. The theoretical properties of graph-based tests such as these are explored in Bhattacharya \cite{bhattacharya2019general}. In a low-dimensional setting, we use simulation studies to demonstrate that the multivariate AUGUST test has comparable power to current methods. 

Finally, in Section \ref{sec:nba}, we apply our method to NBA data on throw distance and angle from the net. Do shots follow a different distribution than misses? How about throws early in the game versus late in the game? In addition to answering these questions, we use AUGUST to create graphical data representations addressing why the null was rejected in each case. Finally, with the help of the multivariate AUGUST test, we revisit the NBA shooting data in the context of joint distributions.

\section{Main insights}
\label{sec:insights}

\subsection{CDF transformation}
Given independent samples $\{\bm{X}_i\}_{i = 1}^m$ and $\{\bm{Y}_i\}_{i = 1}^n$, where $\bm{X}_i\sim G$ and $\bm{Y}_i \sim F$, recall that we are interested in testing
\begin{align*}
    H_0: F = G \text{   versus   } H_a: F\neq G.
\end{align*}
For our purposes, we will assume that $F$ and $G$ are absolutely continuous functions.

To illustrate the main idea, imagine a one-sample setting. We want to test whether or not $\bm{X}_i \sim F$, with $F$ known. Under the null hypothesis $\bm{X}_i \sim F$, it is a well-known result that the transformed variables $\{F(\bm{X}_i):i\in [m]\}$ follow a Uniform$(0, 1)$ distribution. When $\bm{X}$ does not follow $F$, the collection $\{F(\bm{X}_i):i\in [m]\}$ is not uniform. As a result, we can reduce our one-sample test to a test of the uniformity of $\{F(\bm{X}_i):i\in [m]\}$. Moreover, examining \textit{how} the collection $\{F(\bm{X}_i):i\in [m]\}$ fails to be uniform tells us why $F$ does not fit the distribution of $\bm{X}$.

In the two-sample setting, the same intuition holds true: we might construct transformed variables that are nearly uniform in $[0, 1]$ when $F = G$, and that are not uniform otherwise. When the distributions of the two samples are different, the way that uniformity fails should be informative.

Given the fact that the CDF-transformed variables $\{G(\bm{X}_i):i\in [m]\}$ follow a uniform distribution, an intuitive choice would be $\{\hat{F}_{\bm{Y}}(\bm{X}_i): i \in [m]\}$, where $\hat{F}_{\bm{Y}}$ is the empirical CDF of $\bm{Y}$:
\begin{align*}
\hat{F}_{\bm{Y}}(t) = \frac{1}{n}\sum_{i = 1}^{n} I(\bm{Y}_i \leq t).
\end{align*}
The BET framework introduced in Zhang \cite{zhang2019bet} gives us a way to test $\{\hat{F}_{\bm{Y}}(\bm{X}_i): i \in [m]\}$ for uniformity up to a given binary depth $d$, which is equivalent to testing multinomial uniformity over dyadic fractions $\{\frac{1}{2^d},\dots , 1\}$. In particular, we can define a vector $\bm{P}$ of length $2^d$ such that, for $1 \leq i \leq 2^d$,
\begin{align*}
\bm{P}_i = \frac{\#\bigg\{k: \hat{F}_{\bm{Y}}(\bm{X}_k) \in \left[\frac{i-1}{2^d}, \frac{i}{2^d}\right)\bigg\}}{m}.
\end{align*}
Then the vector of symmetry statistics is given by $\bm{S} = \mathbf{H}_{2^d}\bm{P}$, where $\mathbf{H}_{2^d}$ is the Hadamard matrix of size $2^d$ according to Sylvester's construction. In particular, we can restrict our attention to $\bm{S}_{-1}$, since the first coordinate of $\bm{S}$ is always equal to $\sum_{i = 1}^{2^d} \bm{P}_i = 1$. As shown in Zhang \cite{zhang2019bet}, $\bm{S}_{-1}$ is a sufficient statistic for uniformity in the one sample setting, and the BET test based on $\bm{S}_{-1}$ achieves the minimax rate in sample size required for power against a wide variety of alternatives. 

We can think of $\bm{S}_{-1}$ in a signal-processing context: the Hadamard transform maps the vector of cell probabilities $\bm{P}$ in the physical domain to the vector of symmetries $\bm{S}_{-1}$ in the frequency domain. This transformation is advantageous since, in the one sample setting, the entries of $\bm{S}_{-1}$ have mean zero and are pairwise uncorrelated under the null. As a result, fluctuations of $\bm{S}_{-1}$ away from $\bm{0}_{2^d-1}$ unambiguously support the alternative, and the coordinates of $\bm{S}_{-1}$ are interpretable as orthogonal signals of nonuniformity. Moreover, the vector $\bm{P}$ always satisfies $\sum_{i = 1}^{2^d} \bm{P}_i = 1$, meaning that the mass of $\bm{P}$ is constrained to a $(2^d-1)$-dimensional hyperplane in $\mathbb{R}^{2^d}$. In contrast, the vector $\bm{S}_{-1}$ is non-degenerate and summarizes the same information about non-uniformity with greater efficiency.

To clarify this procedure, we provide a concrete example. Consider the case $d = 2$, and suppose our calculation for $\bm{S} = \mathbf{H}_{4}\bm{P}$ can be explicitly written
\begin{align*}
\begin{pmatrix} 1.00 \\ 0.00 \\ 0.50 \\ -0.10 \end{pmatrix} = \begin{pmatrix} 1 & 1 & 1 & 1 \\ 1 & -1 & 1 & -1 \\ 1 & 1 & -1 & -1 \\ 1 & -1 & -1 & 1\end{pmatrix} \begin{pmatrix} 0.35 \\ 0.40 \\ 0.15 \\ 0.10 \end{pmatrix}.
\end{align*}
Note that the first symmetry statistic, $\bm{S}_1 = 1$, is constant and not diagnostic for asymmetry. Of the other statistics, $\bm{S}_3 = 0.5$ is largest in absolute value. This indicates that the greatest imbalance comes from the row $(1, 1, -1, -1)$, which compares the fraction of points in the first half of $[0, 1]$ to the fraction of points in the last half.

One possible choice of test statistic is the quantity $S = \norm{\bm{S}_{-1}}_2^2$. A test based on $S$ is essentially a $\chi^2$ test and has decent power at detecting $F \neq G$. However, it turns out that we can achieve much higher power by constructing $\bm{P}$ a bit differently.

\subsection{An ``Augmented'' CDF}
\label{ssn:augmented}
A problem with directly using the ECDF of $\bm{Y}$ as a transformation for $\bm{X}$ is dependence between the transformed variables $\{\hat{F}_{\bm{Y}}(\bm{X}_i), i \in [m]\}$. While each variable marginally follows a discrete uniform distribution on $\{0, \frac{1}{n}, \dots, 1\}$ under the null, the joint distribution of $\hat{F}_{\bm{Y}}(\bm{X}_1)$ and $\hat{F}_{\bm{Y}}(\bm{X}_2)$ has more mass along the diagonal $\{(0, 0), (\frac{1}{n}, \frac{1}{n}), \dots, (1, 1)\}$. Informally, this is because the events $\{\hat{F}_{\bm{Y}}(\bm{X}_1) = \frac{k}{n}\}$ and $\{\hat{F}_{\bm{Y}}(\bm{X}_2) = \frac{k}{n}\}$ are both more likely to occur when the distance between the order statistics $\bm{Y}_{(k)}$ and $\bm{Y}_{(k+1)}$ is large. Due to this correlation, the symmetry statistics of $\{\hat{F}_{\bm{Y}}(\bm{X}_i), i \in [m]\}$ tend to ``swing'' more heavily in one direction or the other, increasing variance under the null and negatively affecting power.

An unrealistic way to resolve the dependence would be to obtain an entirely different $\bm{Y}$ sample for each transformed variable $\hat{F}_{\bm{Y}}(\bm{X}_i)$. Instead, since we are only interested in the uniformity of $\{\hat{F}_{\bm{Y}}(\bm{X}_i), i \in [m]\}$ up to binary depth $d$, we can decrease the dependence by computing ECDF transformations $\hat{F}_{\bm{Y}^*}(\bm{X}_i)$ based on a small, random subsample $\bm{Y}^*$ of size $r$ from $\bm{Y}$. The following discussion makes this process explicit.

Let $\bm{Y}^*$ be a random subsample from $\bm{Y}$ of size $r = 2^{d+1}-1$. For any $x\in\mathbb{R}$, let $p_k^{\bm{Y}}(x)$ be the probability, conditional on $\bm{Y}$, that either $2k-2$ or $2k-1$ elements of $\bm{Y}^*$ are less than or equal to $x$. It turns out that the probabilities $p_k^{\bm{Y}}(x)$ are essentially hypergeometric and simple to compute:
\begin{align*}
p_k^{\bm{Y}}(x) = \frac{\binom{\#\{i:\bm{Y}_i \leq x\}}{2k-2}\binom{\#\{i:\bm{Y}_i > x\}}{2^{d+1}-1 - (2k-2)}}{\binom{n}{2^{d+1}-1}} + \frac{\binom{\#\{i:\bm{Y}_i \leq x\}}{2k-1}\binom{\#\{i:\bm{Y}_i > x\}}{2^{d+1}-1 - (2k-1)}}{\binom{n}{2^{d+1}-1}}.
\end{align*}
Using the function $p_k^{\bm{Y}}(\cdot)$, we define $\bm{P}_x$ to be the vector of length $2^d$ such that, for each coordinate $k$,
\begin{align*}
\bm{P}_{x, k} = p_{k}^{\bm{Y}}(x), \text{ for } 1 \leq k \leq 2^d.
\end{align*}
Note that $\hat{F}_{\bm{Y}^*}(x) \in \left[\frac{k-1}{2^d}, \frac{k}{2^d}\right)$ exactly when $2k-2$ or $2k-1$ subsampled elements in $\bm{Y}^*$ are less than or equal to $x$. Therefore, we could equally say that 
\begin{align*}
\bm{P}_{x, k} & = P\left(\hat{F}_{\bm{Y}^*}(x) \in \left[\frac{k-1}{2^d}, \frac{k}{2^d}\right)\Bigg|\bm{Y}\right), \text{ for } 1 \leq k \leq 2^d.
\end{align*}
It is in precisely this sense that $\bm{P}_x$ can be considered an ``augmented'' CDF: instead of mapping $x$ to a single value in the unit interval, $x\mapsto \bm{P}_x$ maps $x$ to a distribution. Moreover, this characterization explains the choice of subsample size $r = 2^{d+1}-1$. Any $r$ satisfying $r = 2^{q}-1$, $q \geq d$, guarantees that the discrete random variable $\hat{F}_{\bm{Y}^*}(x)$ has the same number of point masses inside every interval of the form $\left[\frac{k-1}{2^d}, \frac{k}{2^d}\right)$. The specific choice of $q = d+1$ has been found to work best empirically.

To collect information about every $\bm{X}_i$, we define the vector $\bm{P}_{\bm{X}}$, now with a vector subscript, by the average of all $\bm{P}_{\bm{X}_i}$:
\begin{align*}
\bm{P}_{\bm{X}} = \frac{1}{m}\sum_{i = 1}^m \bm{P}_{\bm{X}_i}.
\end{align*}
Given that the formula for $p_k^{\bm{Y}}(x)$ is computed from hypergeometric probabilities, we refer the coordinates of $\bm{P}_{\bm{X}}$ as \textit{hypergeometric cell probabilities}. Just as we expect the distribution of the ECDF-transformed variables $\{\hat{F}_{\bm{Y}}(\bm{X}_i): i \in [m]\}$ to be uniform under the null, we expect the mass of $\bm{P}_{\bm{X}}$ to be nearly uniform over its coordinates. The vector of symmetry statistics $\bm{S}_{\bm{X}} = (\mathbf{H}_{2^d}\bm{P}_{\bm{X}})_{-1}$ quantifies non-uniformity in $\bm{P}_{\bm{X}}$.

Notably, the cell probabilities in $\bm{P}_{\bm{x}}$ are computed in reference to a resampling procedure but without actually resampling. As the discussion above suggests, these probabilities could indeed be approximated by a bootstrap procedure: take many subsamples $\bm{Y}^*$ of size $2^{d+1}-1$ from $\bm{Y}$, compute $\hat{F}_{\bm{Y}^*}(x)$ each time, and bin the results as cell counts at intervals of $1/2^d$. The exact cell probabilities $\bm{P}_{x}$ derived above are the limiting cell probabilities of this bootstrap procedure as the number of resamples tends to infinity. The following theorem makes this result explicit.

\begin{theorem}
\label{thm:resamp}
Here, let $\bm{Y}$ be a fixed vector of observed data, and let $x$ be some fixed number. Consider the following bootstrap method for computing the vector $\bm{P}^*_{x}$ using $K$ subsamples from $\bm{Y}$.
\begin{enumerate}
\item Take bootstrap resamples $\bm{Y}^*_k$ of size $2^{d+1}-1$ from $\bm{Y}$ without replacement, for resamples $1 \leq k \leq K$.
\item Compute $\hat{F}_{\bm{Y}^*_k}(x)$, for resamples $1 \leq k \leq K$.
\item Set $\bm{P}^*_{x, i} = \#\left\{k:\hat{F}_{\bm{Y}^*_k}(x) \in \left[\frac{i-1}{2^d}, \frac{i}{2^d}\right)\right\}/K$, for coordinates $1 \leq i \leq 2^d$.\\

It follows that 
\begin{align*}
P\left(\lim_{K \rightarrow \infty} \bm{P}^*_x = \bm{P}_x\right) = 1,
\end{align*}
where the probability is taken over the randomness of the resampling.
\end{enumerate}
\end{theorem}

Theorem \ref{thm:resamp} shows that the hypergeometric cell probabilities are equivalent to the limiting cell probabilities of a certain bootstrap procedure. Effectively, one could say that actual resampling is a valid way to approximate $\bm{P}_{\bm{X}}$. In practice, and in the current context of univariate data, it is much faster to directly compute the limiting hypergeometric probabilities.

\subsection{Choice of statistic}

Recall the steps to computing $\bm{S}_{\bm{X}}$:
\begin{align*}
p_k^{\bm{Y}}(x) & = \frac{\binom{\#\{i:\bm{Y}_i \leq x\}}{2k-2}\binom{\#\{i:\bm{Y}_i > x\}}{2^{d+1}-1 - (2k-2)}}{\binom{n}{2^{d+1}-1}} + \frac{\binom{\#\{i:\bm{Y}_i \leq x\}}{2k-1}\binom{\#\{i:\bm{Y}_i > x\}}{2^{d+1}-1 - (2k-1)}}{\binom{n}{2^{d+1}-1}}\\
\bm{P}_{x, k} & = p_{k}^{\bm{Y}}(x), \text{ for coordinate } k = 1 \dots 2^d\\
\bm{P}_{\bm{X}} & = \frac{1}{m}\sum_{i = 1}^m \bm{P}_{\bm{X}_i}\\
\bm{S}_{\bm{X}} & = (\mathbf{H}_{2^d}\bm{P}_{\bm{X}})_{-1}.
\end{align*}
The entries of $\bm{P}_{\bm{X}}$ sum to $1$ and are expected to be roughly uniform under the null. The entries of $\bm{S}_{\bm{X}}$ quantify non-uniformity in $\bm{P}_{\bm{X}}$ and correspond to symmetries encoded in the rows of $\mathbf{H}_{2^d}$. Moreover, $\bm{S}_{\bm{X}}$ is a non-degenerate random vector.

With $\bm{S}_{\bm{Y}}$ defined analogously to $\bm{S}_{\bm{X}}$, we propose the statistic $S = -\bm{S}_{\bm{X}}^T\bm{S}_{\bm{Y}}$. First, this choice of statistic has the advantage of treating the $\bm{X}$ and $\bm{Y}$ samples symmetrically. This is desirable because neither sample is assumed to have privileged status, so it would be counterintuitive for the value of $S$ to change when the roles of $\bm{X}$ and $\bm{Y}$ are switched. In addition, this statistic is a continuous function of the concatenated vector $(\bm{S}_{\bm{X}}, \bm{S}_{\bm{Y}})^T$. In Theorem \ref{thm:null}, we give an asymptotic normality result for $(\bm{S}_{\bm{X}}, \bm{S}_{\bm{Y}})^T$, meaning that the asymptotic distribution of $S$ is accessible. The primary insight leading to this result is the fact that the concatenated vector $(\bm{S}_{\bm{X}}, \bm{S}_{\bm{Y}})^T$ can be written as a two-sample $U$-statistic, which is the subject of Theorem \ref{thm:kernel}. Going forward, we will use $\text{AUGUST}(\bm{X}, \bm{Y}, d)$ to denote the test based on $S$.

The negative sign in $-\bm{S}_{\bm{X}}^T\bm{S}_{\bm{Y}}$ arises from the fact that $\bm{S}_{\bm{X}}$ and $\bm{S}_{\bm{Y}}$ are negatively correlated under the alternative, and we want the critical values of $S$ to be positive. The proposition below gives intuition for this negative correlation in the context of a location shift.
\begin{proposition}
Suppose our $\bm{X}$ and $\bm{Y}$ data are such that $\max_i\{\bm{X}_i\} < \min_j\{\bm{Y}_j\}$. Then $\cos (\theta) = -(2^d - 1)^{-1}$, where $\theta$ is the angle between $\bm{S}_{\bm{X}}$ and $\bm{S}_{\bm{Y}}$ as vectors in $\mathbb{R}^{2^d-1}$.
\end{proposition}

In general, we could say the following: if $\bm{X}$ is to the left of $\bm{Y}$, then $\bm{Y}$ is to the right of $\bm{X}$, and the symmetry statistic detecting left/right imbalance will be positive in $\bm{S}_{\bm{X}}$ and negative in $\bm{S}_{\bm{Y}}$. This negative correlation holds true for all symmetry statistics, so we include the negative sign in $-\bm{S}_{\bm{X}}^T\bm{S}_{\bm{Y}}$ so that this product is large in the positive direction.

It is important to note that $\bm{S}_{\bm{X}}$ and $\bm{S}_{\bm{Y}}$ can be negatively correlated under the null, as well. However, $\norm{\bm{S}_{\bm{X}}}_2$ and $\norm{\bm{S}_{\bm{Y}}}_2$ are larger under the alternative than under the null, meaning that $-\bm{S}_{\bm{X}}^T\bm{S}_{\bm{Y}}$ is still larger under the alternative. See Theorems \ref{thm:null} and \ref{thm:alt} for additional exploration of the theoretical properties of $\bm{S}_{\bm{X}}$ and $\bm{S}_{\bm{Y}}$.

\subsection{Interpreting the results}
\label{ssn:interpret}
Suppose we have real $\bm{X}$ and $\bm{Y}$ data, and we wish to test the distributional equality of our samples. Before performing the AUGUST test, we must choose some resolution $d$ -- this decision determines the scale on which AUGUST will be sensitive. For example, $d = 1$ is sensitive only to mean/location shift, as $\tilde{\mathbf{H}}_{2} = \begin{pmatrix} 1 & -1 \end{pmatrix}$ has one row, comparing left/right side probabilities. At $d = 2$, sensitivity to scale emerges, coming from the row $\begin{pmatrix} 1 & -1 & -1 & 1 \end{pmatrix}$ in $\tilde{\mathbf{H}}_{4}$. In practice, $d = 3$ should be sufficient for global distributional differences, which existing ECDF-based methods can detect. In Zhang \cite{zhang2021beauty}, it is shown that a depth of $d = 3$ is sufficient for a symmetry statistic-based test of independence to outperform both distance correlation and $F$-test, which are known to be optimal, in detecting correlation in bivariate normal distributions.

Higher depths $d > 3$ are additionally sensitive to local information -- this can be useful for alternatives that are extremely close in the Kolmogorov-Smirnov metric but have densities that are bounded apart in the uniform norm. As one example, we may have $\bm{X}$ sampled from Uniform$(0, 1)$ and $\bm{Y}$ sampled from a high frequency square wave distribution with the same support.

Given some choice of $d$, suppose we calculate $\bm{P}_{\bm{X}}, \bm{P}_{\bm{Y}}$, as well as $\bm{S}_{\bm{X}} = (\mathbf{H}_{2^d}\bm{P}_{\bm{X}})_{-1}$, and $\bm{S}_{\bm{Y}} = (\mathbf{H}_{2^d}\bm{P}_{\bm{Y}})_{-1}$ as specified before. The AUGUST test based on $S$ rejects the null, claiming that $\bm{X}$ and $\bm{Y}$ come from different distributions. How can we use the AUGUST test to interpret this rejection?

We can consider $\bm{Y}$ as our \textit{reference sample}, meaning that we will make statements about how points of $\bm{X}$ fall relative to the distribution of $\bm{Y}$. In this case, looking at the entries of $\bm{S}_{\bm{X}}$ next to the matrix $\mathbf{H}_{2^d}$ tells us what we want to know. Each entry in the vector $\bm{S}_{\bm{X}}$ specifies the non-uniformity of $\bm{P}_{\bm{X}}$ with respect to a row of $\mathbf{H}_{2^d}$. In particular, the largest entries of $\bm{S}_{\bm{X}}$ in absolute value tell us the sources of greatest asymmetry in $\bm{P}_{\bm{X}}$.

For a concrete example, we let $d = 3$. Given $\bm{X}$ and $\bm{Y}$ data, suppose that $\mathbf{H}_{8}\bm{P}_{\bm{X}}$ is explicitly computed to be
\begin{align*}
\begin{pmatrix} 1.00 \\ 0.00 \\ -0.10 \\ 0.02 \\ -0.02 \\ -0.02 \\ -0.08 \\ 0.00 \end{pmatrix} = \begin{pmatrix} 1 & 1 & 1 & 1 & 1 & 1 & 1 & 1 \\ 1 & -1 & 1 & -1 & 1 & -1 & 1 & -1 \\ 1 & 1 & -1 & -1 & 1 & 1 & -1 & -1 \\ 1 & -1 & -1 & 1 & 1 & -1 & -1 & 1 \\ 1 & 1 & 1 & 1 & -1 & -1 & -1 & -1 \\ 1 & -1 & 1 & -1 & -1 & 1 & -1 & 1 \\ 1 & 1 & -1 & -1 & -1 & -1 & 1 & 1 \\ 1 & -1 & -1 & 1 & -1 & 1 & 1 & -1 \end{pmatrix}\begin{pmatrix} 0.10 \\ 0.10 \\ 0.14 \\ 0.15 \\ 0.13 \\ 0.12 \\ 0.13 \\ 0.13 \end{pmatrix}.
\end{align*}

Recall that $\bm{S}_{\bm{X}}$ consists of all but the first coordinate of the vector on the left. In this case, the vector $\bm{S}_{\bm{X}}$ has two notable entries, which decompose the non-uniformity of $\bm{P}_{\bm{X}}$ into two orthogonal signals. The largest entry of $\bm{S}_{\bm{X}}$ in absolute value is $-0.10$, corresponding to the third row of $\mathbf{H}_{8}$:
\begin{align*}
\begin{pmatrix} 1 & 1 & -1 & -1 & 1 & 1 & -1 & -1 \end{pmatrix}.
\end{align*}
We can interpret this correspondence in the following way: the distribution of $\bm{X}$ has a coarse Venetian blind pattern relative to $\bm{Y}$. The second largest entry of $\bm{S}_{\bm{X}}$ in absolute value is $-0.08$, due to the seventh row of $\mathbf{H}_{8}$:
\begin{align*}
\begin{pmatrix} 1 & 1 & -1 & -1 & -1 & -1 & 1 & 1 \end{pmatrix}.
\end{align*}
From this, we see that the $\bm{X}$ points are centrally concentrated relative to the points of $\bm{Y}$. We would expect the interquartile region of $\bm{Y}$ to contain over half of the points of $\bm{X}$.

Now, suppose we wish to visualize the largest imbalance recorded in $\bm{S}_{\bm{X}}$. In the example above, the largest component of $\bm{S}_{\bm{X}}$ is $-0.10$. Inspecting the corresponding row of $\mathbf{H}_8$, it follows that the third, fourth, seventh, and eighth coordinates of $\bm{P}_{\bm{X}}$ comprise more than half of the total mass in $\bm{P}_{\bm{X}}$.

Let $R_1,\dots, R_8$ be real intervals such that $1/2^d = 1/8$ of the $\bm{Y}$ sample is contained in each $R_i$. These eight intervals correspond to the cells of $\bm{P}_{\bm{X}}$: if $\bm{P}_{\bm{X}, i}$ is large (small), we would expect $R_i$ to contain more (less) than $1/8$ of the $\bm{X}$ sample. In the context of the example above, the regions $R_3, R_4, R_7$, and $R_8$ together contain more than half of the points of $\bm{X}$ but exactly half of the points of $\bm{Y}$. This imbalance reflects the largest asymmetry in $\bm{S}_{\bm{X}}$, and in the context of testing, can be thought of as the primary reason for rejection of the null.

\begin{figure}
\includegraphics[scale = .55]{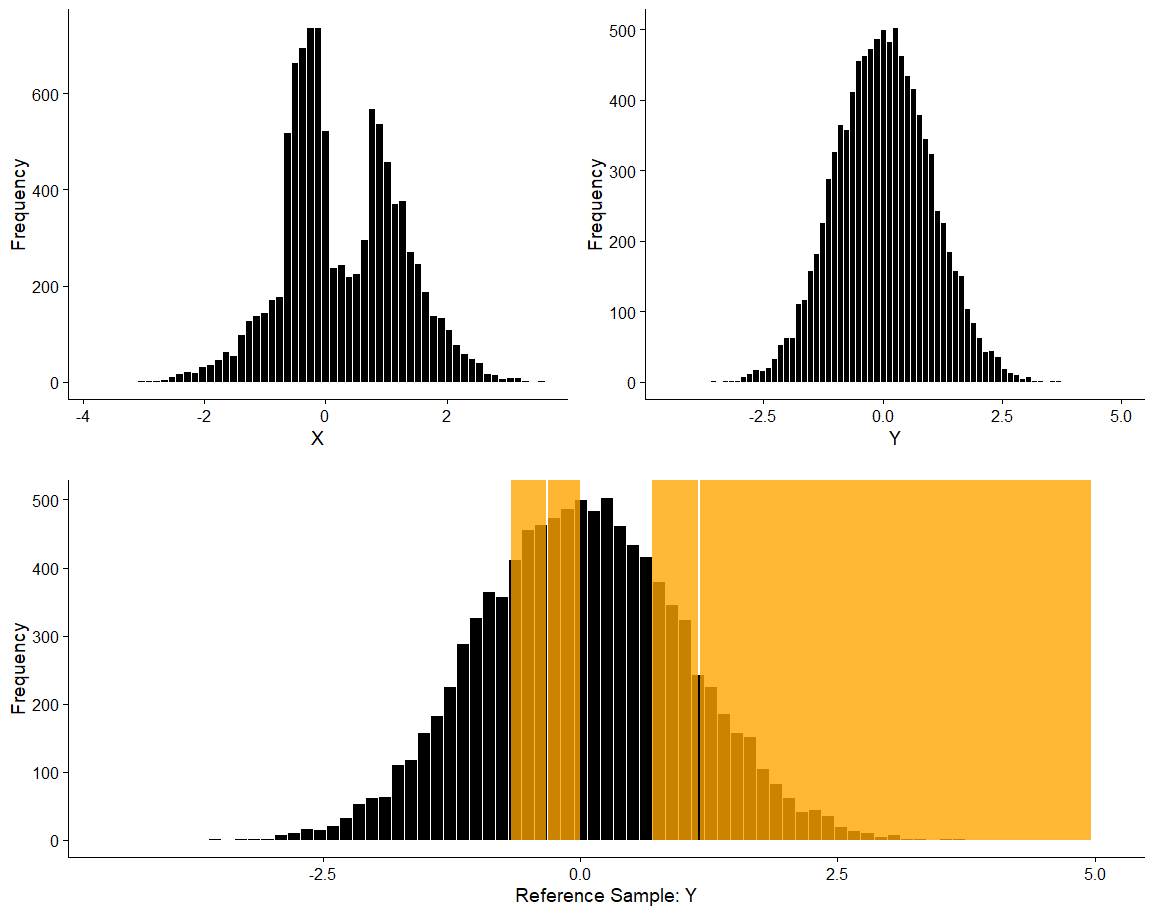}
\caption{Visualization of a symmetry statistic using simulated data. In the concrete example of Section \ref{ssn:interpret}, the largest asymmetry in $\bm{S}_{\bm{X}}$ indicates that $\bm{X}$ has a coarse Venetian blind pattern relative to the reference sample $\bm{Y}$. The yellow rectangles above represent this pattern, as shaded regions contain an excess of $\bm{X}$ points relative to the plotted $\bm{Y}$ sample. As these rectangles represent the largest symmetry statistic in $\bm{S}_{\bm{X}}$, this particular imbalance is interpretable as the primary reason for rejection of the null.}
\label{fig:ex}
\end{figure}

In Figure \ref{fig:ex}, we visualize simulated $\bm{X}$ and $\bm{Y}$ data whose largest asymmetry corresponds to the regions $R_3, R_4, R_7$, and $R_8$. In Section $\ref{sec:nba}$, we use this style of visualization on NBA shooting data.

\section{Method}
\label{sec:method}

\subsection{Algorithms for the AUGUST statistic}

Algorithms \ref{alg:augmentedcdf} and \ref{alg:AUGUST} formalize the steps to the AUGUST test outlined in earlier sections. In terms of the notation from earlier, Algorithm \ref{alg:augmentedcdf} computes the augmented CDF vector $\bm{P}_x^{\bm{V}}$, and Algorithm \ref{alg:AUGUST} performs a complete test using the statistic $S = -\bm{S}_{\bm{X}}^T\bm{S}_{\bm{Y}}$.

\begin{algorithm}
    \caption{AugmentedCDF$(x,\bm{V},d)$}
    \label{alg:augmentedcdf}
    \begin{algorithmic}[1]
        \STATE Initialize zero vector $\bm{P}_x$ of length $2^d$
        \STATE $N = \text{length}(\bm{V}_i)$
        \STATE $n = 2^{d+1} - 1$
        \STATE $K = \#\{i: \bm{V}_i \leq x\}$
        \FOR{$i = 1$ to $2^d$}
            \STATE $k = 2i - 2$
            \STATE $\bm{P}_{x, i} = \frac{\binom{K}{k}\binom{N-K}{n - k}}{\binom{N}{n}} + \frac{\binom{K}{k + 1}\binom{N-K}{n - k - 1}}{\binom{N}{n}}$
        \ENDFOR
        \STATE Return $\bm{P}_x$
    \end{algorithmic}
\end{algorithm}

\begin{algorithm}
    \caption{AUGUST$(\bm{X},\bm{Y},d)$}
    \label{alg:AUGUST}
    \begin{algorithmic}[1]
        \STATE Initialize zero vectors $\bm{P}_{\bm{X}}$, $\bm{P}_{\bm{Y}}$ of length $2^d$
        \FOR{$i = 1$ to $\text{length}(\bm{X})$}
            \STATE $\bm{P}_{\bm{X}} = \bm{P}_{\bm{X}} + \text{AugmentedCDF}(\bm{X}_i, \bm{Y}, d)$

        \ENDFOR
        \FOR{$i = 1$ to $\text{length}(\bm{Y})$}
            \STATE $\bm{P}_{\bm{Y}} = \bm{P}_{\bm{Y}} + \text{AugmentedCDF}(\bm{Y}_i, \bm{X}, d)$

        \ENDFOR
        \STATE Assign $\bm{P}_{\bm{X}} = \bm{P}_{\bm{X}}/\text{length}(\bm{X})$ and $\bm{P}_{\bm{Y}} = \bm{P}_{\bm{Y}}/\text{length}(\bm{Y})$
        \STATE Assign $\bm{S}_{\bm{X}} = (\mathbf{H}_{2^d}\bm{P}_{\bm{X}})_{-1}$ and $\bm{S}_{\bm{Y}} = (\mathbf{H}_{2^d}\bm{P}_{\bm{Y}})_{-1}$
        \STATE Compute the statistic $S = -\bm{S}_{\bm{X}}^T\bm{S}_{\bm{Y}}$
        \STATE Reject when $S$ is large
    \end{algorithmic}
\end{algorithm}

Recall that our two samples $\bm{X}$ and $\bm{Y}$ have sizes $m$ and $n$, respectively. Treating $d$ as a constant, Algorithm \ref{alg:AUGUST} requires $O(mn)$ elementary operations. This is due to the line $K = \#\{i: \bm{V}_i \leq x\}$ in the function AugmentedCDF$(x,\bm{V},d)$, which necessitates iterating over all entries of $\bm{V}$ each time that AugmentedCDF$(x,\bm{V},d)$ is called. In the function AUGUST, the vectors $\bm{X}$ and $\bm{Y}$ are passed into AugmentedCDF$(x,\bm{V},d)$ as the argument $\bm{V}$ a total of $n$ and $m$ times, respectively.

However, there exists a more efficient implementation of AUGUST for instances when $n$ and $m$ are both large. By first sorting the concatenated $\bm{X}$ and $\bm{Y}$ samples, it is possible to reduce the running time to $O((m+n)\log(m+n))$ operations. The improved algorithm, dubbed AUGUST+, is recorded as Algorithm \ref{alg:AUGUST+}, and the running time of AUGUST+ is recorded in Theorem \ref{thm:runningtime}.

\begin{algorithm}
    \caption{AUGUST+$(\bm{X},\bm{Y},d)$}
    \label{alg:AUGUST+}
    \begin{algorithmic}[1]
        \STATE Define $m = \text{length}(\bm{X})$, $n = \text{length}(\bm{Y})$, and $r = 2^{d+1}-1$
        \STATE Initialize empty matrix $\mathbf{M}$ of dimension $2\times (m+n)$
        \STATE Assign the first row of $\mathbf{M}$ to the concatenated vector $(\bm{X}^T, \bm{Y}^T)$
        \STATE Assign the second row of $\mathbf{M}$ to a row vector with $m$ entries equal to $1$ followed by $n$ entries equal to $0$
        \STATE Sort the columns of $\mathbf{M}$ ascending by the entries in the first row of $\mathbf{M}$
        \STATE Initialize integers $c_x, c_y = 0$ and vectors $\bm{P}_{\bm{X}}, \bm{P}_{\bm{Y}} = \mathbf{0}_{2^d}$
        \FOR{$i = 1$ to $(m+n)$}
            \IF{$\mathbf{M}_{2, i} = 1$}
                \STATE $c_x = c_x + 1$
                \FOR{$j = 1$ to $2^d$}
                    \STATE $k = 2j-2$
                    \STATE $\bm{P}_{\bm{X}, j} = \bm{P}_{\bm{X}, j} + \displaystyle\frac{\binom{c_y}{k}\binom{n-c_y}{r - k}}{\binom{n}{r}} + \frac{\binom{c_y}{k + 1}\binom{n-c_y}{r - k - 1}}{\binom{n}{r}}$
                \ENDFOR
            \ELSE
                \STATE $c_y = c_y + 1$
                \FOR{$j = 1$ to $2^d$}
                    \STATE $k = 2j-2$
                    \STATE $\bm{P}_{\bm{Y}, j} = \bm{P}_{\bm{Y}, j} + \displaystyle\frac{\binom{c_x}{k}\binom{m-c_x}{r - k}}{\binom{n}{r}} + \frac{\binom{c_x}{k + 1}\binom{m-c_x}{r - k - 1}}{\binom{m}{r}}$
                \ENDFOR
            \ENDIF
        \ENDFOR
        \STATE Assign $\bm{P}_{\bm{X}} = \bm{P}_{\bm{X}}/m$ and $\bm{P}_{\bm{Y}} = \bm{P}_{\bm{Y}}/n$
        \STATE Assign $\bm{S}_{\bm{X}} = (\mathbf{H}_{2^d}\bm{P}_{\bm{X}})_{-1}$ and $\bm{S}_{\bm{Y}} = (\mathbf{H}_{2^d}\bm{P}_{\bm{Y}})_{-1}$
        \STATE Compute the statistic $S = -\bm{S}_{\bm{X}}^T\bm{S}_{\bm{Y}}$
        \STATE Reject when $S$ is large
    \end{algorithmic}
\end{algorithm}

\begin{theorem}
\label{thm:runningtime}
AUGUST+$(\bm{X}, \bm{Y}, d)$ runs in $O((m+n)\log(m+n))$ time.
\end{theorem}

In Table \ref{tab:run}, we provide simulation results confirming the conclusion of Theorem \ref{thm:runningtime}. The second column of Table \ref{tab:run} records the running time of Kolmogorov-Smirnov for each sample size, and the third column indicates the running time of the AUGUST+ test, another $O(N\log N)$ algorithm. This implementation of the Kolmogorov-Smirnov test comes from the \texttt{twosamples} R package and uses a pre-compiled C++ function to compute the test statistic, with a default of 2000 permutations. The AUGUST+ algorithm used here is implemented solely in R and uses a known critical value from theory across all tests. The final column of Table \ref{tab:run} gives the ratio of running times for the two tests, demonstrating that AUGUST+ is significantly faster than a permutation-based Kolmogorov-Smirnov test for every sample size considered.

\begin{table}
\label{tab:run}
\begin{center}
\begin{tabular}{ r | r | r | r}
    $N$ & Kolmogorov-Smirnov (sec)& AUGUST+ (sec)& Ratio (KS/AUG+)\\
    \hline
    $10^2$ &  0.06 & 0.01 & 4.98\\
    $10^3$ & 0.27 & 0.07 & 3.93\\
    $10^4$ & 2.49 &  0.60 & 4.13\\
    $10^5$ & 30.36 &  5.95 & 5.11\\
    $10^6$ & 416.47 &  42.21 & 9.87
\end{tabular}
\end{center}
\caption{Comparison of running times for Kolmogorov-Smirnov and AUGUST+. Critical values for KS are computed using the default 2000 bootstrapped resamples, while critical values for the AUGUST test use the convergence of $S$ via theory. In every case, AUGUST+ is several times faster than KS, and the result of Theorem \ref{thm:runningtime} is supported.}
\end{table}

\subsection{Multivariate extensions}
\label{sec:multi}

Above, we have taken $\bm{X}$ and $\bm{Y}$ to be univariate iid samples from some distributions $G$ and $F$. It turns out that we can extend the univariate AUGUST test to the problem of multivariate two-sample testing. Suppose that $\bm{X}$ and $\bm{Y}$ are iid samples from some continuous distributions $G$ and $F$, with each $\bm{X}_i$ and $\bm{Y}_j$ in $\mathbb{R}^k$, for some $k \geq 2$ and $1 \leq i \leq m$, $1 \leq j \leq n$. In order to use the AUGUST test, our goal is to transform the multivariate $\bm{X}$ and $\bm{Y}$ data into univariate samples $\tilde{\bm{X}}$ and $\tilde{\bm{Y}}$. The transformed $\tilde{\bm{X}}$ and $\tilde{\bm{Y}}$ should be equal in distribution exactly when the multivariate null hypothesis $H_0: F = G$ is true. The exact form of this transformation determines the geometric interpretation of the cell probabilities computed via AUGUST. One technique to achieve elliptical cells could be appropriately named \textit{mutual Mahalanobis distance}. 

Given a mean $\bm{\mu}\in\mathbb{R}^k$ and invertible $k\times k$ covariance matrix $\mathbf{\Sigma}$, recall that the Mahalanobis distance of $\bm{x}\in\mathbb{R}^k$ from $\bm{\mu}$ with respect to $\mathbf{\Sigma}$ is given by
\begin{align*}
MD(\bm{x}; \bm{\mu}, \mathbf{\Sigma}) = \sqrt{(\bm{x} - \bm{\mu})^T\mathbf{\Sigma}^{-1}(\bm{x} - \bm{\mu})}.
\end{align*}
Now, let $\hat{\bm{\mu}}_{\bm{X}}$ and $\hat{\mathbf{\Sigma}}_{\bm{X}}$ be the sample mean and sample covariance matrix of $\bm{X}$. Consider the transformed collections
\begin{align*}
\tilde{\bm{X}}^{(\bm{X})} & = \left\{ MD(\bm{X}_i; \hat{\bm{\mu}}_{\bm{X}}, \hat{\mathbf{\Sigma}}_{\bm{X}}): 1 \leq i \leq m \right\} \\
\tilde{\bm{Y}}^{(\bm{X})} & = \left\{ MD(\bm{Y}_j; \hat{\bm{\mu}}_{\bm{X}}, \hat{\mathbf{\Sigma}}_{\bm{X}}): 1 \leq j \leq n \right\},
\end{align*}
where the superscript $(\bm{X})$ indicates that means and covariances are estimated using the $\bm{X}$ sample. If $\bm{X}$ and $\bm{Y}$ come from the same multivariate distribution, then the collections $\tilde{\bm{X}}^{(\bm{X})}$ and $\tilde{\bm{Y}}^{(\bm{X})}$ should have similar univariate distributions. As a result, given some depth $d$, we can compute the AUGUST statistic for the samples $\tilde{\bm{X}}^{(\bm{X})}$ and $\tilde{\bm{Y}}^{(\bm{X})}$ in order to test for the distributional equality of $\bm{X}$ and $\bm{Y}$.

Recall that the AUGUST test can be thought of as testing for regions of imbalance between $\bm{X}$ and $\bm{Y}$ -- imbalances in distribution appear as non-uniformity in the vector of cell probabilities. Under this Mahalanobis distance transformation, the cells used by AUGUST correspond to nested elliptical rings centered on $\hat{\bm{\mu}}_{\bm{X}}$. 

As in the univariate case, it is desirable for the test statistic to be invariant to the transposition of $\bm{X}$ and $\bm{Y}$. To achieve this, we can use the test statistic
\begin{align*}
S_{multi} = \max\left(AUGUST(\tilde{\bm{X}}^{(\bm{X})}, \tilde{\bm{Y}}^{(\bm{X})}, d), AUGUST(\tilde{\bm{X}}^{(\bm{Y})}, \tilde{\bm{Y}}^{(\bm{Y})}, d)\right),
\end{align*}
wherein we use \textit{both} possible Mahalanobis distance transformations for $\bm{X}$ and $\bm{Y}$, compute two AUGUST statistics, and take the maximum. As we show in Section \ref{sec:performance}, a depth of $d = 2$ is sufficiently large to detect common multivariate alternatives.

\section{Theoretical results}
\label{sec:theoretical}
\subsection{General $U$-statistic theory}

Here, we present theory necessary for specifying the asymptotic distribution of the test statistic $S$ for the univariate AUGUST test. Many of the arguments in this subsection imitate the technique and notation of Asymptotic Statistics by A. W. van der Vaart \cite{van2000asymptotic}, extending the $U$-statistic theory of Chapters 11 and 12 to a multivariate kernel. We begin by stating two key lemmas:

\begin{lemma}[Orthogonality of the projection]
\label{lem:orthogofproj}
Let $\bm{U}\in\mathbb{R}^p$ be a random vector, and let $\{W_i\}_{i = 1}^N$ be a collection of $N$ independent observations. Define the projection $\hat{\bm{U}} = E\bm{U} + \sum_{i = 1}^N E(\bm{U} - E\bm{U} | W_i)$. Then
\begin{align*}
E(\bm{U} - \hat{\bm{U}})\hat{\bm{U}}^T = \mathbf{0}_{p\times p}
\end{align*}
\end{lemma}

\begin{lemma}[Closeness of the projection]
\label{lem:closenessofproj}
Let $\{W_i\}_{i = 1}^\infty$ be an independent collection of random variables. Let $\{\bm{U}_N\}_{N = 1}^{\infty}$ be a sequence of non-degenerate random vectors of length $p$. For each $N$, define the projection $\hat{\bm{U}}_N = E\bm{U}_N + \sum_{i = 1}^N E(\bm{U}_N - E\bm{U}_N | W_i)$. Let $\mathbf{\Sigma}_{1, N} = \text{Cov}(\bm{U}_N)$ and $\mathbf{\Sigma}_{2, N} = \text{Cov}(\hat{\bm{U}}_N)$. If $\mathbf{\Sigma}_{1, N}\mathbf{\Sigma}_{2, N}^{-1} \rightarrow I$ as $N \rightarrow \infty$, then
\begin{align*}
\mathbf{\Sigma}_{1, N}^{-\frac{1}{2}}\left(\bm{U}_{N} - E\bm{U}_{N}\right) - \mathbf{\Sigma}_{2, N}^{-\frac{1}{2}}\left(\hat{\bm{U}}_{N} - E\hat{\bm{U}}_{N}\right)  \overset{p}{\to}  0.
\end{align*}
\end{lemma}

Using Lemmas \ref{lem:orthogofproj} and \ref{lem:closenessofproj}, one can prove asymptotic normality of $U$-statistics. Suppose we have independent samples $\{X_i\}_{i = 1}^m$ and $\{Y_j\}_{j = 1}^n$, where $\bm{X}_i \sim G$ and $\bm{Y}_j \sim F$. Recall that a two-sample $U$-statistic based on $\bm{X}$ and $\bm{Y}$ has form
\begin{align*}
\bm{U} = \frac{1}{\binom{m}{r}\binom{n}{s}}\sum_{\alpha}\sum_{\beta}\bm{k}(\bm{X}_{\alpha_1}, \dots, \bm{X}_{\alpha_r}, \bm{Y}_{\beta_1}, \dots, \bm{Y}_{\beta_s})
\end{align*}
where $\bm{k}:\mathbb{R}^r\times\mathbb{R}^s \rightarrow \mathbb{R}^p$ is called the kernel corresponding to $\bm{U}$. We impose the restriction that $\bm{k}$ is symmetric in its first $r$ and last $s$ coordinates. Notation-wise, the index $\alpha$ is a combination of length $r$ from the set $\{1, \dots, m\}$, and the outer sum is taken over all such combinations. The index $\beta$ and inner sum are analogous. We can think of $\bm{U}$ as an unbiased estimator of $$\bm{\theta} := E\bm{k}(\bm{X}_1,\dots,\bm{X}_{r}, \bm{Y}_1, \dots, \bm{Y}_{s}).$$

\begin{theorem}
\label{thm:u}
Let $N = n + m$, and assume that $n, m \rightarrow \infty$ in such a way that $m/N \rightarrow \lambda$ for some $\lambda \in (0, 1)$. Define the cross-covariance matrices
\begin{align*}
\xi_{i, j} = \text{Cov}\Bigg(&\bm{k}(\bm{X}_1, \dots,\bm{X}_{r}, \bm{Y}_1, \dots,  \bm{Y}_{s}),\\
& \bm{k}(\bm{X}_1, \dots, \bm{X}_i, \bm{X}_{i+1}', \dots, \bm{X}_{r}', \bm{Y}_1, \dots, \bm{Y}_{j}, \bm{Y}_{j+1}', \dots, \bm{Y}_{s}')\Bigg).
\end{align*} 
If $\mathbf{\Sigma} = r^2\xi_{1, 0}/\lambda + s^2\xi_{0, 1}/(1-\lambda)$ is invertible, then
\begin{align*}
\sqrt{N}\left(\bm{U} - \bm{\theta}\right) \xrightarrow{d} N(\bm{0}, \mathbf{\Sigma}).
\end{align*}
\end{theorem}

The proof of Theorem \ref{thm:u} relies on invertibility of the limiting covariance matrix. It turns out that as long as this matrix has rank at least one, we still achieve asymptotic normality of $\bm{U}$, albeit possibly to a degenerate distribution.
\begin{theorem}
\label{thm:udeg}
If $\mathbf{\Sigma} = r^2\xi_{1, 0}/\lambda + s^2\xi_{0, 1}/(1-\lambda)$ has rank $q \geq 1$, then
\begin{align*}
\sqrt{N}\left(\bm{U} - \bm{\theta}\right) \xrightarrow{d} N(\bm{0}, \mathbf{\Sigma}).
\end{align*}
\end{theorem}

In general, it can be difficult to prove or disprove the invertibility of the limiting covariance matrix $\mathbf{\Sigma}$. In light of Theorem \ref{thm:udeg}, we can say that $\sqrt{N}\left(\bm{U} - \bm{\theta}\right)$ converges in distribution to some multivariate normal as long as $\mathbf{\Sigma}$ is not identically zero. In particular, this is true whenever some coordinate of $\bm{k}_{1, 0}(\bm{X}_1)$ or $\bm{k}_{0, 1}(\bm{Y}_1)$ (defined in the proof of Theorem \ref{thm:u}) has nonzero variance, which is often trivial to show.

\subsection{Writing $S$ as a function of a $U$-statistic}

Now that we have general results for $U$-statistics, we need to relate our test statistic $S = -\bm{S}_{\bm{X}}^T\bm{S}_{\bm{Y}}$ to a $U$-statistic in some way. In this subsection, we work toward showing that the concatenated vector of symmetry statistics $\begin{pmatrix} \bm{S}_{\bm{X}} \\ \bm{S}_{\bm{Y}} \end{pmatrix}$ is in fact a $U$-statistic.

Let $d\in\mathbb{N}$ be the fixed binary depth, and let the function $\bm{h}:\mathbb{R}\times\mathbb{R}^{2^{d+1}-1} \rightarrow \mathbb{R}^{2^d}$ be given by
\[ \bm{h}_k(x, \bm{y}) =  \begin{cases} 
      1 & \text{if } \#\{j: \bm{y}_j \leq x\} = 2k - 2 \text{ or } 2k - 1\\
      0 & \text{otherwise.}
   \end{cases}
\]
The following key lemma explains how $\bm{P}_{\bm{X}}$ can be expressed using $\bm{h}$.
\begin{lemma}
\label{lem:h}
With $\bm{h}$ as defined above, it holds that
\begin{align*}
\frac{1}{\binom{n}{2^{d+1}-1}} \sum_{\beta} \bm{h}\left(x, \bm{Y}_{\beta_1}, \bm{Y}_{\beta_2}, \dots, \bm{Y}_{\beta_{2^{d+1}-1}}\right) = \text{AugmentedCDF}(x, \bm{Y}, d).
\end{align*}
Consequently,
\begin{align*}
\frac{1}{m\binom{n}{2^{d+1}-1}} \sum_i\sum_{\beta} \bm{h}\left(\bm{X}_i, \bm{Y}_{\beta_1}, \bm{Y}_{\beta_2}, \dots, \bm{Y}_{\beta_{2^{d+1}-1}}\right) = \bm{P}_{\bm{X}}.
\end{align*}
\end{lemma}

For intuition on the above result, we can look to the classic urn model. Consider an urn with $n$ balls: one red ball for each $\bm{Y}_i \leq x$, and one black ball for each $\bm{Y}_i > x$. Subsampling $2^{d+1}-1$ points from $\bm{Y}$ is equivalent to drawing $2^{d+1}-1$ balls from the urn. In this case, the $k$th coordinate of $\bm{h}$ is an indicator of the event that exactly $2k-2$ or $2k-1$ red balls were drawn. By averaging $\bm{h}_k$ over every possible combination of balls from the urn, we compute the probability of this event. Computing the probability this way is inefficient compared to the obvious hypergeometric approach, but this form ultimately allows us to write $\begin{pmatrix} \bm{S}_{\bm{X}} \\ \bm{S}_{\bm{Y}} \end{pmatrix}$ as a $U$-statistic.

\begin{theorem}
\label{thm:kernel}
There exists a kernel function
$$\bm{k}: \mathbb{R}^{2^{d+1}-1}\times\mathbb{R}^{2^{d+1}-1} \rightarrow \mathbb{R}^{2^d-1}\times\mathbb{R}^{2^d-1}$$
such that
\begin{align*}
& \frac{1}{\binom{m}{2^{d+1}-1}\binom{n}{2^{d+1}-1}}\sum_\alpha\sum_\beta \bm{k}(\bm{X}_{\alpha_1},\dots, \bm{X}_{\alpha_{2^{d+1}-1}}, \bm{Y}_{\beta_1},\dots, \bm{Y}_{\beta_{2^{d+1}-1}}) = \begin{pmatrix} \bm{S}_{\bm{X}} \\ \bm{S}_{\bm{Y}} \end{pmatrix}.
\end{align*}
\end{theorem}

In summary, we have shown that the concatenated vector of symmetry statistics $\begin{pmatrix} \bm{S}_{\bm{X}} \\ \bm{S}_{\bm{Y}} \end{pmatrix}$ is a vector-valued, two-sample $U$-statistic. This result opens the door to asymptotic results for $S = -\bm{S}_{\bm{X}}^T\bm{S}_{\bm{Y}}$ under both the null and alternative.

\subsection{Asymptotic normality}

First, we address the asymptotic distribution of $S$ under the null. 
\begin{theorem}
\label{thm:null}
Suppose that we have univariate iid observations $\{\bm{X}_i\}_{i = 1}^m$ and $\{\bm{Y}_j\}_{j = 1}^n$ under the null. Let $N = n + m$, and assume that $n, m \rightarrow \infty$ in such a way that $m/N \rightarrow \lambda$ for some $\lambda \in (0, 1)$. Then 
\begin{align*}
\sqrt{N}\begin{pmatrix} \bm{S}_{\bm{X}} \\ \bm{S}_{\bm{Y}} \end{pmatrix} \xrightarrow{d} N(\bm{0}, \mathbf{\Sigma}).
\end{align*}
Defining the cross-covariance matrices
\begin{align*}
\xi_{i, j} = \text{Cov}\Bigg(&\bm{k}(\bm{X}_1, \dots,\bm{X}_{2^{d+1}-1}, \bm{Y}_1, \dots,  \bm{Y}_{2^{d+1}-1}),\\
& \bm{k}(\bm{X}_1, \dots, \bm{X}_i, \bm{X}_{i+1}', \dots, \bm{X}_{2^{d+1}-1}', \bm{Y}_1, \dots, \bm{Y}_{j}, \bm{Y}_{j+1}', \dots, \bm{Y}_{2^{d+1}-1}')\Bigg),
\end{align*}
the auto-covariance matrix $\mathbf{\Sigma}$ is given by
\begin{align*}
\mathbf{\Sigma} = (2^{d+1}-1)^2\left(\xi_{1, 0}/\lambda + \xi_{0, 1}/(1-\lambda)\right).
\end{align*}
\end{theorem}

Under the null, Theorem \ref{thm:udeg} and the continuous mapping theorem specify the asymptotic distribution of $S = -\bm{S}_{\bm{X}}^T\bm{S}_{\bm{Y}}$ as an inner product of central, correlated normal random vectors. Under the alternative, one would expect $S$ to be noncentral in some sense, where the amount of noncentrality is dictated by the way in which $F \neq G$. This turns out to be the case, as the next theorem indicates.

First, we provide some definitions used in the next theorem statement and proof. For each $k\in \{1, \dots, 2^d\}$, define the function $p_k^F:\mathbb{R}\rightarrow [0, 1]$ by
\begin{align*}
p_k^F(x) & = \binom{2^{d+1}-1}{2k-2}F(x)^{2k-2}(1-F(x))^{2^{d+1}-1 - (2k-2)} \\
& + \binom{2^{d+1}-1}{2k-1}F(x)^{2k-1}(1-F(x))^{2^{d+1}-1 - (2k-1)},
\end{align*}
and similarly define $p_k^G:\mathbb{R}\rightarrow [0, 1]$ by
\begin{align*}
p_k^G(x) & = \binom{2^{d+1}-1}{2k-2}G(x)^{2k-2}(1-G(x))^{2^{d+1}-1 - (2k-2)} \\
& + \binom{2^{d+1}-1}{2k-1}G(x)^{2k-1}(1-G(x))^{2^{d+1}-1 - (2k-1)}.\\
\end{align*}
These functions can be thought of as theoretical analogs of the probabilities $p_k^{\bm{Y}}(x)$ and $p_k^{\bm{X}}(x)$ from Section \ref{ssn:augmented}. Further, define the quantities
\begin{align*}
p_k^{F:G} = \int p_k^F(x)dG(x)
\end{align*}
and
\begin{align*}
p_k^{G:F} = \int p_k^G(x)dF(x).
\end{align*}

\begin{theorem}
\label{thm:alt}
Suppose that we have univariate, independent observations $\{\bm{X}_i\}_{i = 1}^m$ and $\{\bm{Y}_j\}_{j = 1}^n$, where $\bm{X}_i \sim G$ and $\bm{Y}_j \sim F$. Let $N = n + m$, and assume that $n, m \rightarrow \infty$ in such a way that $m/N \rightarrow \lambda$ for some $\lambda \in (0, 1)$. Then  
\begin{align*}
\sqrt{N}\left(\begin{pmatrix} \bm{S}_{\bm{X}} \\ \bm{S}_{\bm{Y}} \end{pmatrix} - \bm{\mu}\right) \xrightarrow{d} N(\bm{0}, \mathbf{\Sigma}),
\end{align*}
where $\bm{\mu}$ is given by
\begin{align*}
\bm{\mu} = \begin{pmatrix} \tilde{\mathbf{H}}_{2^d} & \mathbf{0}_{(2^d-1)\times 2^d} \\ \mathbf{0}_{(2^d-1)\times 2^d} & \tilde{\mathbf{H}}_{2^d} \end{pmatrix}\begin{pmatrix} p_1^{F:G}\\ \vdots \\ p_{2^d}^{F:G} \\[6pt] p_1^{G:F}\\ \vdots \\ p_{2^d}^{G:F}\end{pmatrix}.
\end{align*}
Defining the cross-covariance matrices
\begin{align*}
\xi_{i, j} = \text{Cov}\Bigg(&\bm{k}(\bm{X}_1, \dots,\bm{X}_{2^{d+1}-1}, \bm{Y}_1, \dots,  \bm{Y}_{2^{d+1}-1}),\\
& \bm{k}(\bm{X}_1, \dots, \bm{X}_i, \bm{X}_{i+1}', \dots, \bm{X}_{2^{d+1}-1}', \bm{Y}_1, \dots, \bm{Y}_{j}, \bm{Y}_{j+1}', \dots, \bm{Y}_{2^{d+1}-1}')\Bigg),
\end{align*}
where expectations are taken under the alternative, the auto-covariance matrix $\mathbf{\Sigma}$ is given by
\begin{align*}
\mathbf{\Sigma} = (2^{d+1}-1)^2\left(\xi_{1, 0}/\lambda + \xi_{0, 1}/(1-\lambda)\right).
\end{align*}
\end{theorem}

As one consequence of the above result, given distributions $F \neq G$, it is possible to compute the limit in probability of the symmetry statistics $\begin{pmatrix} \bm{S}_{\bm{X}} \\ \bm{S}_{\bm{Y}} \end{pmatrix}$. This limit $\bm{\mu}$ encodes asymmetry at the population level, analogous to how $\begin{pmatrix} \bm{S}_{\bm{X}} \\ \bm{S}_{\bm{Y}} \end{pmatrix}$ encodes asymmetry between the finite samples $\bm{X}$ and $\bm{Y}$. Moreover, using this theorem, one can efficiently simulate the AUGUST statistic under the alternative, making it easy to benchmark AUGUST against a predetermined $F \neq G$ in large samples. In applications that require an \textit{a priori} power analysis, this approach can simplify the process of determining the sample size necessary for detecting a given effect.

\section{Empirical performance}
\label{sec:performance}

\subsection{Univariate performance}
\label{sec:uniper}

\begin{figure}
\includegraphics[scale = .54]{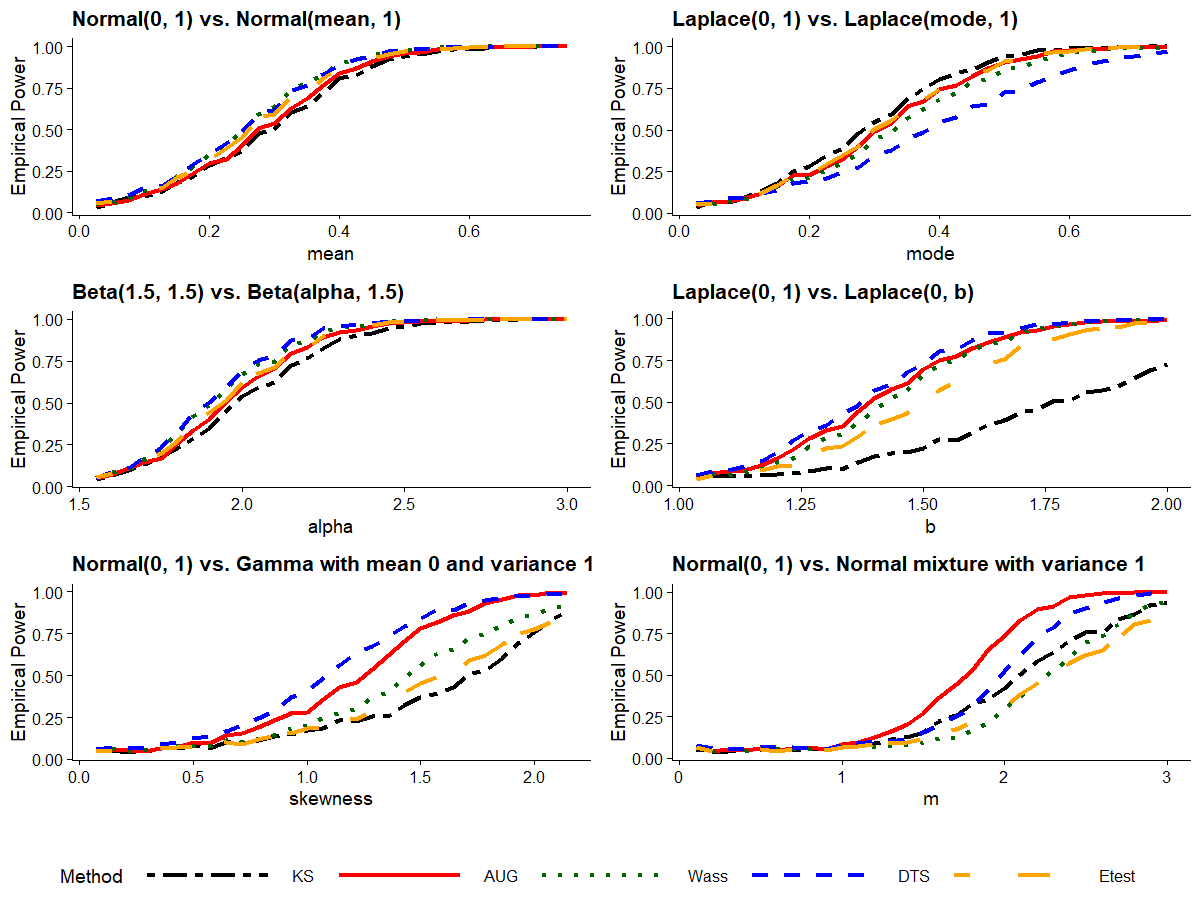}
\caption{Comparison of power for non-parametric univariate two-sample tests. We consider the energy distance test (Etest), Kolmogorov-Smirnov (KS), Wasserstein (Wass), DTS, and AUGUST at depth $d = 3$, all with a cutoff of $\alpha = 0.05$ and sample size $n = m = 128$. No test uniformly outperforms all others, though AUGUST is robust against the range of alternatives and never performs worst. All distributions considered are straightforward except perhaps the normal mixture on the bottom right. The parameter $m$ tracks the separation between the mixed Gaussians, and as $m \rightarrow \infty$, the alternative distribution approaches a Rademacher. The AUGUST test dominates all other tests on this alternative.}
\label{fig:compare}
\end{figure}

In this section, we compare AUGUST to a sampling of other non-parametric two-sample tests: Kolmogorov-Smirnov, Wasserstein, and the recent DTS. We also consider the energy distance test, described in Sz\'ekely and Rizzo \cite{szekely2013energy}. For these simulations, we use a sample size of $n = m = 128$, and for the AUGUST test, we set a depth of $d = 3$. Simulation results are graphed in Figure \ref{fig:compare}. 

The first row of Figure \ref{fig:compare} consists of normal and Laplace location alternatives -- situations where differences in the first distributional moment are most diagnostic. Center left, we have a symmetric beta vs. asymmetric beta alternative. While this does not constitute a pure location shift, differences in first moment are most pronounced. Center right, we include a Laplace scale family. The bottom row of Figure 1 focuses on families with identical first and second moments: normal vs. mean-centered gamma on the bottom left, and normal vs. symmetric normal mixture on the bottom right.

For the location alternatives, the power of each method depends on the shape of the distribution. DTS, Wasserstein, and energy distance perform slightly better than AUGUST for normal and beta distributions, and AUGUST in turn outperforms Kolmogorov-Smirnov. In contrast, for a Laplace location shift, Kolmogorov-Smirnov outperforms every test, with AUGUST in second place and DTS last. For the Laplace scale family, Kolmogorov-Smirnov performs badly, with DTS and AUGUST leading.

On the more complicated alternatives, Wasserstein, KS, and energy distance suffer power loss relative to their earlier performance. DTS slightly outperforms AUGUST on the gamma skewness family, while AUGUST outperforms all other tests at detecting normal vs. normal mixture.

The most important lesson is this: no single test performs best in all situations. Even for simple alternatives such as location families, the precise shape of the distribution is highly influential as to the tests' relative performance. Indeed, the performance rankings of DTS, Wasserstein, energy distance, and KS in the Laplace location trials are exactly \textit{reversed} compared to the normal location trials. Notably, the AUGUST test never performs worst out of the methods examined. We theorize that this is because the symmetry statistics $\bm{S}_{\bm{X}}$ and $\bm{S}_{\bm{Y}}$ are weighted equally in every coordinate, meaning that AUGUST is very parsimonious towards the range of potential alternatives. In contrast, the other methods are highly sensitive to location and scale shifts, but they are less robust against more obscure alternatives. The current field of tests may also favor location and scale shifts because these are among the most intuitive families to use for benchmarking. However, this behavior is undesirable in applications that truly require a non-parametric test.

\subsection{Multivariate performance}

\begin{figure}
\includegraphics[scale = .54]{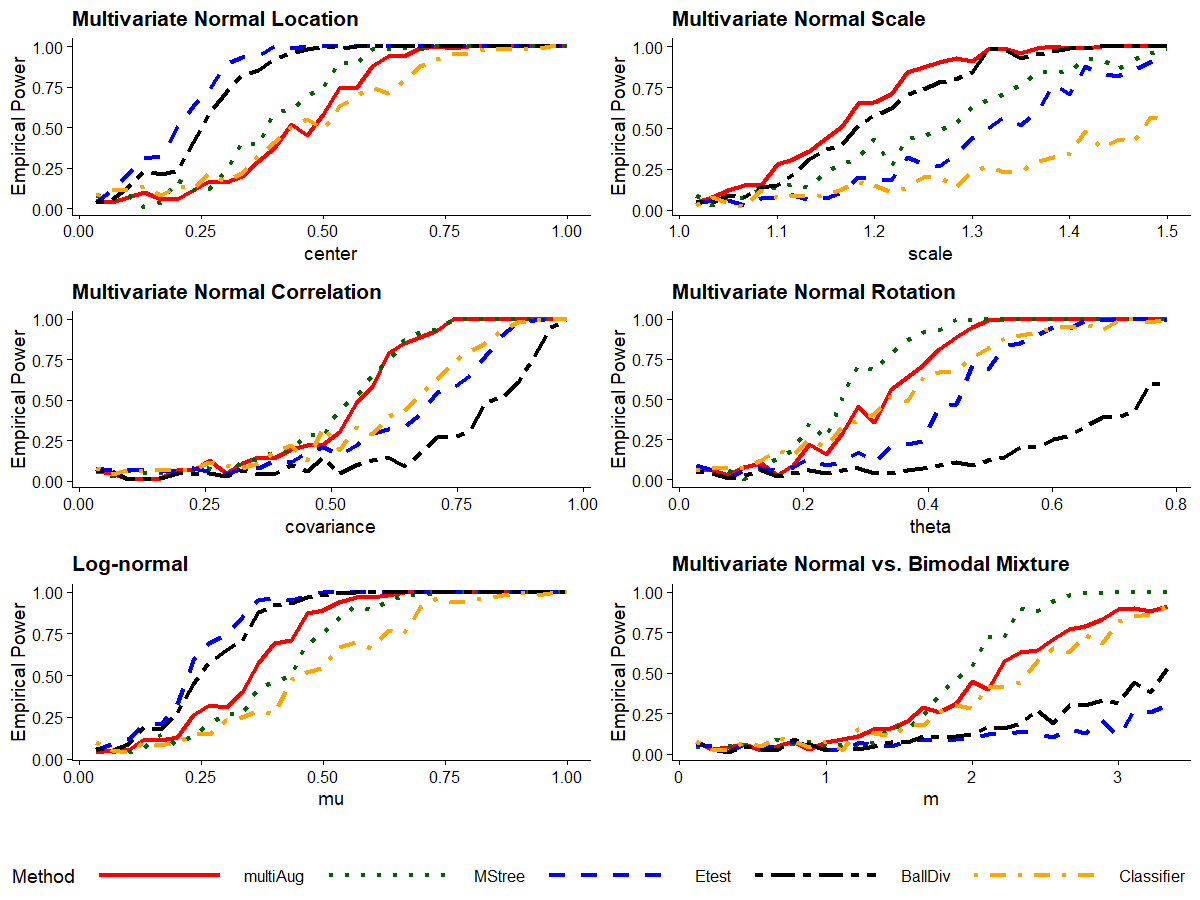}
\caption{Comparison of power for multivariate non-parametric two-sample tests at dimension $k = 2$ and sample size $n = m = 128$. For comparison with AUGUST, we consider the energy distance test of Sz\'ekely and Rizzo \cite{szekely2013energy}, the generalized edge-count method of Chen and Friedman \cite{chen2017new} using 5-minimum spanning trees, the ball divergence test of Pan et. al. \cite{pan2018ball}, and the classifier test of Lopez-Paz and Oquab \cite{lopez2016revisiting}. All tests use a threshold of $\alpha = .05$, and the multivariate AUGUST test is performed at a depth $d = 2$. The performance of AUGUST is comparable to that of existing methods in all circumstances, and AUGUST has superior performance against scale alternatives.}
\label{fig:multicompare}
\end{figure}

In Figure \ref{fig:multicompare}, we compare the mutual Mahalanobis version of AUGUST to some other well-known non-parametric multivariate two-sample tests in a low-dimensional context ($k = 2$). In particular, we again consider the energy distance test of Sz\'ekely and Rizzo \cite{szekely2013energy}, as well as the generalized edge-count method of Chen and Friedman \cite{chen2017new}, the ball divergence test of Pan et. al. \cite{pan2018ball}, and the classifier test of Lopez-Paz and Oquab \cite{lopez2016revisiting}. For the graph-based method, we use a 5-minimum spanning tree based on Euclidean interpoint distance. 

We consider a variety of alternatives -- moving left to right and top to bottom:
\begin{enumerate}
\item $N_2(\bm{0}, \mathbf{I}_2)$ vs. $N_2(\text{center}\times\bm{1}, \mathbf{I}_2)$
\item $N_2(\bm{0}, \mathbf{I}_2)$ vs. $N_2(\bm{0}, \text{scale}\times \mathbf{I}_2)$
\item $N_2\left(\bm{0}, \begin{pmatrix} 1 & 0\\ 0 & 1\end{pmatrix}\right)$ vs. $ N_2\left(\bm{0}, \begin{pmatrix} 1 & \text{cov}\\ \text{cov} & 1\end{pmatrix}\right)$
\item $N_2\left(\bm{0}, \begin{pmatrix} 1 & 0\\ 0 & 9\end{pmatrix}\right)$ vs. $\mathbf{R}_\theta N_2\left(\bm{0}, \begin{pmatrix} 1 & 0\\ 0 & 9\end{pmatrix}\right)$, where $\mathbf{R}_\theta$ is the $2\times 2$ rotation matrix through an angle $\theta$
\item $\exp\left(N_2(\bm{0}, \mathbf{I}_2)\right)$ vs. $\exp\left(N_2(\mu\times\bm{1}, \mathbf{I}_2)\right)$
\item $N_2(\bm{0}, \mathbf{I}_2)$ vs. $(Z, B)$, where $Z\sim N(0, 1)$ and $B$ independently follows the bimodal mixture distribution from Section \ref{sec:uniper}
\end{enumerate}

In Figure \ref{fig:multicompare}, we see that the energy and ball divergence tests dominate the other methods when mean shift is a factor (i.e. in the normal location and log-normal families). On a scale alternative, AUGUST has the best power, with ball divergence at a close second. In contrast, for correlation, rotation, and multimodal alternatives, the edge-count test and AUGUST have superior power, with ball divergence and energy distance coming at or near last place. 

Overall, we can say that AUGUST is robust against a wide range of possible alternatives, and it has particularly high performance against a scale alternative, where it outperforms all other methods considered. We theorize that, in part, this is because some of the other methods rely so heavily on interpoint distances. The scale alternative does not result in good separation between $\bm{X}$ and $\bm{Y}$, meaning that interpoint distances are not as diagnostic as they would be in, say, a location shift.

\section{Studies of NBA shooting data}
\label{sec:nba}
In this section, we demonstrate the interpretability of AUGUST using 2015-2016 NBA play-by-play data. Consider the distributions of throw distances and angles from the net -- are these distributions different for shots and misses? How about for the first two quarters versus the last two quarters? To address these questions, we acquired play-by-play data for the 2015-2016 NBA season. For each throw, the location of the throw was recorded as a pair of $x,y$ coordinates. These coordinates were converted into a distance and angle from the target net, using knowledge of NBA court dimensions.

Four separate AUGUST tests at a depth of $d = 3$ were performed to analyze the distribution of throw distances and angles; data were split according to shots versus misses and early game versus late game. A Bonferroni correction was applied to the resulting $p$-values. At the $\alpha = .05$ level, throw distance and angle follow different distributions for shots versus misses as well as early versus late game.

\begin{figure}
\includegraphics[scale = .6]{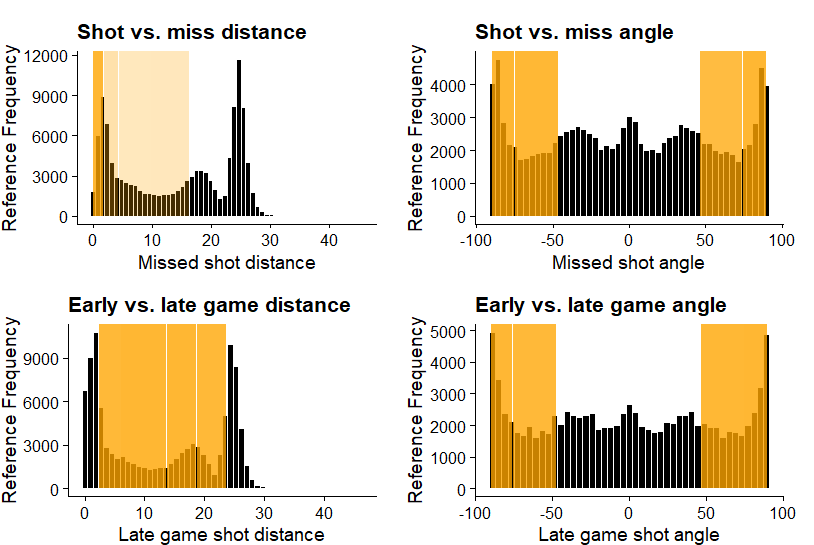}
\caption{Distributional differences in NBA data. Each of the four plots corresponds to a two-sample test. One of the samples from each test is plotted as a histogram -- we can refer to this sample as the reference. Yellow rectangles indicate regions where the reference sample is less prevalent than its counterpart. Since the yellow rectangles correspond to the largest computed symmetry statistic, these plots indicate the primary reason that rejection of the null occurred in each test. In the left column, a peak in shot frequency occurs immediately after the three point line at 23 feet, as intuition would suggest.}
\label{fig:bbfigs}
\end{figure}

To demonstrate the unique interpretability of this test, we provide AUGUST visualizations in Figure \ref{fig:bbfigs} as introduced in Section \ref{ssn:interpret}. Each histogram corresponds to one of the two samples in the test -- this sample is indicated on the $x$-axis. The yellow rectangles overlaid on these histograms illustrate the largest symmetry statistic from the corresponding test. For example, the top left plot corresponds to throw distance for shots versus misses. The histogram records the distribution of missed throw distances, and the yellow bars indicate that successful throws tend to be closer to the net. The width of each bar accounts for $1/2^d = 1/8$ of the sample plotted in the histogram.

Each plot in Figure \ref{fig:bbfigs} yields a specific interpretation as to the greatest distributional imbalance:

\begin{itemize}

\item \textit{Top left:} Compared to misses, successful throws tend to be closer to the net.

\item \textit{Top right:} Successful throws come from the side more often than misses.

\item \textit{Bottom left:} Throws early in the game are more frequently from an intermediate distance than late game throws.

\item \textit{Bottom right:} Relative to the late game, throws early in the game come more frequently from the side.

\end{itemize}

This second bullet point is perhaps most counterintuitive -- conventional wisdom would suggest that throws from in front of the net are more accurate than throws from the sides. This apparent paradox comes from the fact that throws from the sides are typically at a much closer range.

To conclude this section, we again test for equality in distribution of NBA throw distance and angle, now using a multivariate approach. Applying the mutual Mahalanobis distance method of AUGUST with a cutoff of $\alpha = .05$ and depth $d = 2$, we find that the joint distribution of angles and distances differ across shots/misses as well as early/late game, as one would expect given the result of the univariate tests. Interpreting this conclusion is more difficult than in the univariate case due to the way that the Mahalanobis transformation flattens the data into one dimension: ``cells'' in this case correspond to nested elliptical rings centered on the sample means. Constructing an informative visualization for the multivariate setting may be an interesting problem for future work.

\section{Discussion}

Two-sample testing problems arise in a variety of application areas; often the distribution of either sample is unknown. In this paper, we introduce a non-parametric two-sample test dubbed AUGUST, which tests for differences in distribution between two samples up to a predetermined binary depth $d$. This new statistic is distribution-free in finite samples and can be computed in $O(N\log N)$ elementary operations, where $N$ is the total number of observations across both samples. We propose a multivariate extension of AUGUST, allowing for multi-dimensional tests of distributional equality. In addition, we use $U$-statistic theory to specify the asymptotic distribution of the AUGUST statistic, giving the potential for fast $p$-value calculations in a large sample setting. Via simulation studies, we compare the performance of the univariate and multivariate AUGUST tests to that of other well-known non-parametric tests on a variety of distribution families. We find the performance of AUGUST to be comparable to that of the other tests and superior in some cases, such as at detecting unimodality versus bimodality. In order to showcase the interpretability of AUGUST in a real-world setting, we apply our test to NBA shooting data.

This approach admits several directions for future work. Our statistic only uses rank information from the $\bm{X}$ and $\bm{Y}$ samples, disregarding information about distance between observations. While the AUGUST test retains good power and has the benefit of distribution independence in finite samples, it is possible that the power could be further improved by incorporating point distances in some way. 

In a multivariate context, the prototype test of Section \ref{sec:multi} may serve as a useful starting point for future depth-based methods. The current asymptotic theory applies only to the univariate test, meaning that multivariate $p$-value computation requires permutation. Moreover, our present multivariate test is essentially the univariate method in disguise. A future extension of our method to the high-dimensional realm likely calls for a truly multivariate AUGUST-style test.

More broadly speaking, the already difficult problem of non-parametric two-sample testing is even harder in the context of time series. An interpretable, depth-based test may prove useful for the purposes of change point detection.

\section*{Acknowledgements}

The authors thank Hao Chen for valuable comments and suggestions.

\bibliographystyle{plain}
\bibliography{bibliography.bib}

\end{document}